# THE ALGORITHMIC AUTOREGULATION SOFTWARE DEVELOPMENT METHODOLOGY

## AUTORREGULAÇÃO ALGORÍTMICA: A METODOLOGIA DE DESENVOLVIMENTO DE SOFTWARE

(artigo submetido em setembro de 2013)

**Renato Fabbri**
Main Founder, LabMacambira.sf.net (AA)
D. Sc. Candidate. São Carlos Physics Institute – University of São Paulo (IFSC/USP)
fabbri@usp.br

**Vilson Vieira**
Tech Lead, LabMacambira.sf.net (AA)
Graduate student, São Carlos Physics Institute – University of São paulo (IFSC/USP)
vilsonvieira@usp.br

**Danilo Shiga**
Software Engineer, LabMacambira.sf.net (AA)
daniloshiga@gmail.com

**Alexandre Negrão**
Software Engineer, LabMacambira.sf.net (AA)
Undergraduate student, Institute for Computer and Mathematical Sciences, University of São Paulo (ICMC/USP)
bzum07@gmail.com

**Ricardo Fabbri**
Professor Adjunto, Polytechnic Institute – Rio de Janeiro State University (IPRJ/UERJ)
Tech Lead, LabMacambira.sf.net (AA)
rfabbri@iprj.uerj.br

**Daniel Penalva**
Research Scientist, LabMacambira.sf.net (AA)
dkajah@gmail.com

**Marcos Mendonça**
Software Engineer, LabMacambira.sf.net (AA)
marcosm@gmail.com

**Lucas Zambianchi**
Software Engineer, LabMacambira.sf.net (AA)
kamiarc@gmail.com

**GABRIELA SALVADOR THUMÉ**
Software Engineer, LabMacambira.sf.net (AA)
Graduate Student, Institute for Computer and Mathematical Sciences,
University Of São Paulo (ICMC/USP)
GABITHUME@GMAIL.COM

***ABSTRACT***

*We present a new self-regulating methodology for coordinating distributed team work called Algorithmic Autoregulation (AA), based on recent social networking concepts and individual merit. Team members take on an egalitarian role, and stay voluntarily logged into so-called AA sessions for part of their time (e.g. 2 hours per day), during which they create periodical logs — short text sentences — they wish to share about their activity with the team. These logs are publicly aggregated in a Website and are peer-validated after the end of a session, as in code review. A short screencast is ideally recorded at the end of each session to make AA logs more understandable. This methodology has shown to be well-suited for increasing the efficiency of distributed teams working on what is called Global Software Development (GSD), as observed in our experience in actual real-world situations. This efficiency boost is mainly achieved through 1) built-in asynchronous on-demand communication in conjunction with documentation of work products and processes, and 2) reduced need for central management, meetings or time-consuming reports. Hence, the AA methodology legitimizes and facilitates the activities of a distributed software team. It thus enables other entities to have a solid means to fund*

*these activities, allowing for new and concrete business models to emerge for very distributed software development. AA has been proposed, at its core, as a way of sustaining self-replicating hacker initiatives. These claims are discussed in a real case-study of running a distributed free software hacker team called Lab Macambira.*

## RESUMO

O artigo apresenta uma nova metodologia para a coordenação do trabalho de uma equipe dispersa fisicamente chamada Autorregulação algorítmica (AA). A metodologia se baseia em conceitos recentes de redes sociais e mérito individual. Os membros da equipe assumem papéis igualitários e se mantêm logados voluntariamente a sessões de AA por parte do seu tempo (por exemplo, duas horas por dia), criando *logs* periódicos — frases curtas — que desejam compartilhar com os demais envolvidos nas atividades da equipe. Estes *logs* são agregados publicamente em um *website* e são validados pelos pares após o fim da sessão, da mesma forma que se faz na revisão de código. Preferencialmente, um breve *screencast* é gravado ao final de casa sessão para tornar os *logs* de AA mais compreensíveis. Esta metodologia se demonstrou adequada para aumentar a eficiência de equipes dispersas fisicamente trabalhando em projetos de Desenvolvimento de Software Global (GSD), conforme observado em nossa experiência em situações de uso cotidiano. O aumento de eficiência é obtido principalmente por meio de: 1) comunicação assíncrona e sob demanda em conjunto com a documentação dos produtos do trabalho e processos, e 2) necessidade reduzida de gestão centralizada, reuniões ou relatórios que consomem tempo. Assim, a metodologia AA legitima e facilita as atividades de uma equipe de desenvolvimento de software distribuída. Ela possibilita que outras entidades disponham de meios para financiar essas atividades, possibilitando que novos e concretos modelos de negócio se tornem possíveis para desenvolvimentos de software muito distribuídos. A AA foi proposta, em sua essência, como uma forma de possibilitar a auto-replicação de iniciativas de atividade hacker. Estes argumentos são discutidos com base em um estudo de caso real de atuação de uma equipe hacker de software livre distribuído chamada Lab Macambira.

# 1 INTRODUCTION

One of the defining features of modern times is the widening geographical distribution of software teams (LAST, 2003) leading to what is called Global Software Development (GSD) (GERMAN, 2003; FRYER AND GOTHE, 2008; BEGEL, 2008). Paramount examples stem from the free software movement. Projects and institutions such as Mozilla Foundation have several employees, thousands of volunteers and freelance developers distributed across many countries. The same holds for GNOME (GERMAN, 2003), OpenBSD, MySQL or Apache Software Foundation, to cite a few of the most active projects[1]. Their commitment to the public transparency of source code and development processes places them at the global scale of the open Internet. GSD has also seen a growing demand in virtually every other niche of the software industry, even among traditional companies limited to proprietary licensing. This phenomenon is attributed to a variety of factors such as the opportunity to harness a much larger labor pool, the massive globalization of software companies and the search for cheaper production costs (KOMI-SIRVIO, 2005).

Despite the clear advantages of GSD, it is often associated with difficult problems as series of qualitatively new situations arise. For instance, the problem of coordinating and funding free software initiatives on an expressively larger scale than currently practiced is widely held as a tough challenge. Distributed teams are highly heterogeneous, comprising not only volunteers and very experienced developers, but also contractors and freelancers from different backgrounds and cultures. These observations are founded on the factors suggested by Carmel and Agarwal (2001) as main difficulties for GSD: geographic, temporal and cultural differences. In the case of free or open software projects, all such factors are exacerbated, especially given the need to reliably profit in a wildly heterogeneous environment in order to scale up.

Another problem faced by modern software companies and other collectives is the rise of frequent ineffective meetings, which are seldom focused on the particular interest of any attendant. As a result, it has become the norm to take part in too many meetings with open laptops and flashing mobile gadgets, which can be unproductive. Software developers have a valuable creative tendency – they find it enjoyable to code, to be hands-on with their project, to do what they are best at. They despise having to forcibly stop for meetings or to do other bureaucratic activities such as writing lengthy reports to justify their funding (THOMPSON, 2012). In GSD, there is a heavier demand on team coordination, which, in the absence of a proper methodology, can lead to excessive and ineffective on-line meetings and bureaucracy (FRYER AND GOTHE, 2008). Intrinsic geographic, temporal and cultural differences lead to unavoidable issues such as network latency, calling for a different strategy.

To address these matters is the purpose of the AA methodology reported here and the associated software system for coordinating distributed team work. Team members take on an egalitarian role, and stay voluntarily logged in the system for part of their time (e.g. 2 hours per day), during which they log a periodical short text sentence or micrologــ — similar to a ‘tweet’ from Twitter — to sample the status of their activity. Logging is carried out using a series of client UI alternatives: UNIX shell commands, native GUI or Web page, conventional social network posts, or chat messages to a log bot listening to IRC, GTalk, G+, and others. These “microblog sentences” are publicly aggregated and validated by other team members.

[1] The open source network Ohloh has a more complete and constantly updated list of the most active projects on-line at www.ohloh.net.

Through AA, the community has a methodology and an associated system to help implement and validate the activities of a distributed software team. This forms a participation architecture (WEST; O'MAHONY 2008) designed to legitimize financial support for scaling up the activity of distributed development teams. The AA methodology is especially useful for coordinating distributed and decentralized team work, providing effective means to asynchronously update different team members without the need for synchronous unproductive meetings, while ensuring baseline productivity.

A brief overview of current work in GSD methodologies related to AA is presented in section 2. In section 3 the most relevant characteristics of the AA methodology are outlined. In section 4 we report an actual use case of AA for coordinating a team of nine paid developers during the second half of 2011, as well as a broader use case of AA from 2012 to 2014. Section 5 lists overall conclusions and indications of future possibilities for the practical use of AA in other types of teams of software developers or organizations working on non-software distributed activities.

## 2 RELATED WORK

There has been a large amount of research on methodologies to deal with distributed teams of developers. Although this paper focuses on GSD, many of its principles can be brought back down to the conventional setting of smaller teams of developers working at nearly the same place, time zone and with minor cultural differences, depending on the specific context and demands. Moreover, ‘distributed development’ is generally regarded as being global, which is not always true. For instance, AA has been effectively applied to teams whose members live in the same city but work at different timeframes at different locations, see section 4. Even smaller groups of developers working on the same building could use GSD methodologies (or an adapted subset) to their benefit, e.g., to account for different work habits, minimize formal meetings, document work process and history, and so on. A thorough survey of distributed GSD methodologies is beyond the scope of this paper; this section presents but a brief overview.

Various methodologies for GSD were built around the factors that affect distributed teamwork. As proposed by Carmel (1999), these comprise three distances: geographical, cultural and temporal. First, geographical distance handicaps (i) coordination, the act of integrating all the tasks distributed between units (CARMEL AND AGARWAL, 2001); (ii) control, or the process to maintain specific goals, policies or quality levels; and (iii) communication. All those factors are correlated, e.g., a team needs to have clear communication to work on tasks of a specific problem.

Second, cultural distance encompasses differences in organizational and natural culture. Spoken language, individual and ethnic values are common dimensions impacting such distance. Some companies prefer to allocate development units to foreign locations with minimum cultural variance, e.g., an American company may prefer Ireland due to spoken language similarity (CARMEL AND AGARWAL, 2001). Third, temporal distance hampers synchronous communications such as telephone or video conferences. Units of developers working on different time-zones are concerned with managing their agendas in face of temporal dissimilarity.

Targeting geographical distance, Carmel and Agarwal (2001) suggest a strategy to mitigate reliance on synchronous collaboration. Their approach divides the software life-cycle into levels of complexity, each having a degree of collaboration. For example, some developers working on a project with high collaboration demands should use the follow-the-sun approach: when concluding the work day, they pass their work to the team working in another time-zone. Other tactics are suggested by the same author to deal with the three distances, such as separating foreign units of developers in time-zone bands.

Battin *et al.* (2001) propose and discuss their experiments using specific methodologies created for the distributed development centers from Motorola (at the time having 25+ software development centers worldwide). These methodologies included constant communication with critical units, incremental integration and schedules based on time-zones of developers distributed over 6 countries from 3 different continents.

In considering free software projects, similar factors are present and more specialized methodologies arise. German (2003) provides a concise review of the methodologies used by the GNOME project, one of the most active of all free software projects. The manuscript is centered on software architecture. It begins by explaining that GNOME is separated into modules (76 on version 2.4, to be precise) and each module has one maintainer who divides her modules into separate parts within which other developers can work on independent tasks, along other responsibilities. All development is carried out using modern standard free software engineering tools: a bug tracker for bug and issue management, mailing lists and Internet Relay Chat (IRC) for discussion and communication, and a version control system like Git or Mercurial. Periodic (commonly yearly) conferences like GUADEC are held for face-to-face meetings and are hosted in a different location each time, a common practice on other free and open source projects.

Other major free and open source projects employ similar development strategies and tools, with relatively minor variations (REIS AND FORTES, 2003). The Scilab team, who needs to tackle a complex software system related to Matlab, employs similar software engineering tools as GNOME but focuses on Git and code reviews. There is also a yearly periodic conference ScilabTEC in addition to *adhoc* meetings with freelancers. Major development-related modifications to Scilab are proposed by the greater community through Scilab enhancement proposals – SEPs, which are requests for enhancements augmented with design proposals.

We have collaborated with Scilab directly as individual developers, through Google Summer of Code, and through an instance of freelance development funded by the French company Scilab Enterprises to our collective Lab Macambira and universities in Brazil. Freelance development was centered on implemented features – main developers set the desired features to be implemented, and the freelancers stipulate a price to be charged per feature. Some payment was anticipated before the work started, and the other half secured only when the features had been fully implemented. Deadlines were motivated by payment or else set according to milestones for the next release. The development process was otherwise loosely managed and undocumented, so that ambitious goals were avoided due to the risk of no payment or side products of hard work.

Most other major free and open source software such as Blender for 3D modeling, the PureData real-time multimedia programming project and the VXL computer vision libraries, all of which we have helped develop, employ the aforementioned tools except for code review. Similar to GNOME, these software systems are stable enough to be well organized into conceptual modules and file structure, enabling most changes and communication to be localized to module maintainers or past committers. Collective goals and major changes are coordinated directly via e-mail and IRC with the appropriate module maintainer.

The big problem with the above development approaches is that, to this date, free and open source development models have not managed to consistently scale up production levels to match top quality traditional closed-source software in many important niches. Examples include producing professional-grade multimedia applications through the current free software development model such as nonlinear video editors, complex videogames, high throughput real-time video applications, interactive music synthesizers, and scientific visualizers (Matlab, Mathematica and Maple). The simple and direct ways in which the proposed methodology improves upon the aforementioned previous approaches to promote funding and productivity is discussed in section 4. When properly employed in conjunction with existing approaches, the mechanisms underlying AA can help create more viable business models for large-scale free software production, attracting the key parties of sponsors and contributors to projects (SANTOS AND NELSON, 2010; SANTOS *et al.*, 2013). This is confirmed by our experiments.

## 3 THE AA METHODOLOGY

Some of the strategies for GSD mentioned in the previous section are based on complex methodologies, many of which were created for a specific company or software center. This section describes an alternative methodology based on a simple and generally applicable idea: short sessions of focused work periodically logged by a computational tool. Figure 1 summarizes the methodology. It is important to stress that AA is an adaptable methodology that needs to be judiciously customized in practice, at the service of bottom-line team productivity.

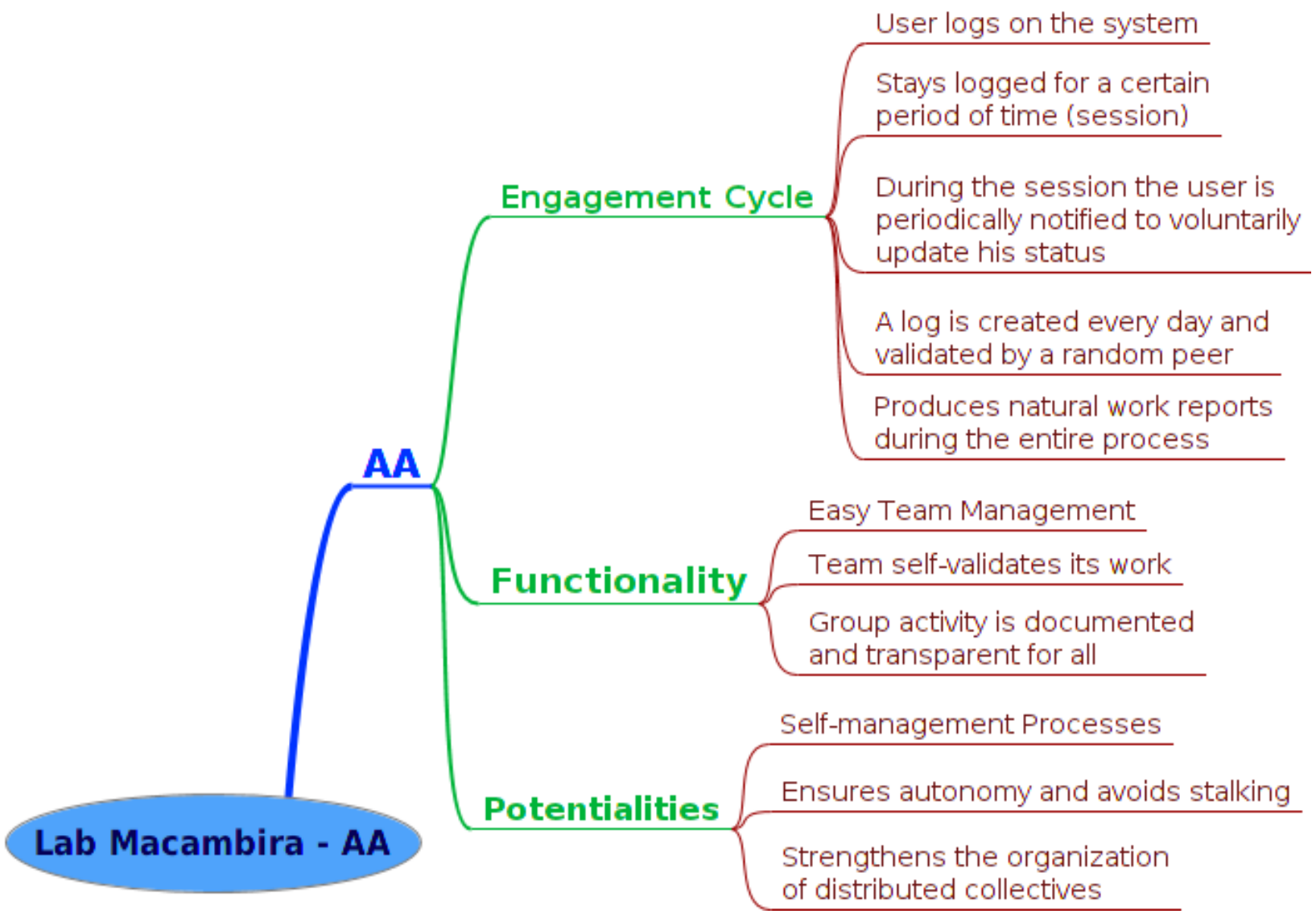

Figure 1: A mind map of the AA methodology: *i. developer engagement cycle* – the usage of AA; *ii. functionality* – the design goals of the system; *iii. potentialities* – end benefits of AA to team context.
Source: the authors.

## 3.1 THE AA SESSION

*"Central to our feelings of awareness is the sensation of the progression of time."*
*– Sir Roger Penrose*

From the developers' perspective, the AA methodology is based on publishing high level individual reports of what they are doing on a specific period of time, quantized into short timeslots and aggregated at various levels. Production status during a timeslot is sampled through so-called micrologs or AA shouts. The timeframe between micrologs, the timeslot size, can range between 5 to 15 minutes in our proposed practice – this can be adjusted depending on what is most convenient for each developer and the team. An AA session is a larger unit of focused continuous work, lasting about 2 hours in our proposed regime, quantized into timeslots. During this time, each developer issues a collection of AA shouts on whatever she wishes to share, normally once per timeslot. Developers have the option to set reminders or AA alerts to show up when it is time to microlog.

The objective of the discretized timeframe and flexible alert scheme is to minimize developer overhead during his AA session and to reduce noise in the published information. The developer can issue meaningful micrologs while staying maximally focused on his code. Each microlog is usually sent directly to an on-line AA server, or stored locally in a temporary database for sending/pushing later on. This enables offline micrologging and periodic alerting.

Each developer optionally records a brief video screencast at the end of her session summarizing what has been done, explaining her goals and challenges in her own words and showcasing her most important results. This is similar to the video logging system in the film Avatar, although it is clear, from our July 2011 Git repositories and online wiki, that we have used this powerful concept in AA independently of its appearance in mass media. Furthermore, screencasting typically captures actual workflow on the computer screen, going beyond videologging in the context of software engineering. When combined with the textual log of the AA session, screencasting renders the final report more understandable by the individual developer himself (increasing self-awareness) and to other people interested in his production (increasing social awareness).

## 3.2 THE AA WEBSITE REPORT

*"Nothing is more important than to see the sources of invention which are, in my opinion, more interesting than the inventions themselves."*

*– Gottfried Leibniz*

All AA reports generated by the development team are continuously sent to a web server and are publicly aggregated on a dashboard website called pAAnel (Figure 2). It is then possible for managers or fellow developers to easily follow the work of any given developer, nearly real-time, reading the small reports or micrologs of what she is working on and how.

It is moreover possible to lookup older sessions to check when certain tasks were carried out, or analyze the comments of the developer about her creative process in tackling a hard problem. Since each AA microlog happens in a short timeslot, the information about what was done – especially *how* it was done – becomes easy to understand, as opposed to a less dynamic report at the end of a session. This is incidentally exploiting the concept of time journals – which also happen to be naturally split up into 15 minute timeslots (BLISS, 1987) – and certain time-management techniques (GOBBO, F. AND VACCARI, M., 2008) on a ***social*** level.

In the current version of the AA server infrastructure, the aggregating website allows the developer to attach a link to her screencast for each worked session. Aggregating screencasts is especially useful when microloging was deliberately rushed, e.g., the developer had to focus on something critical at that moment.

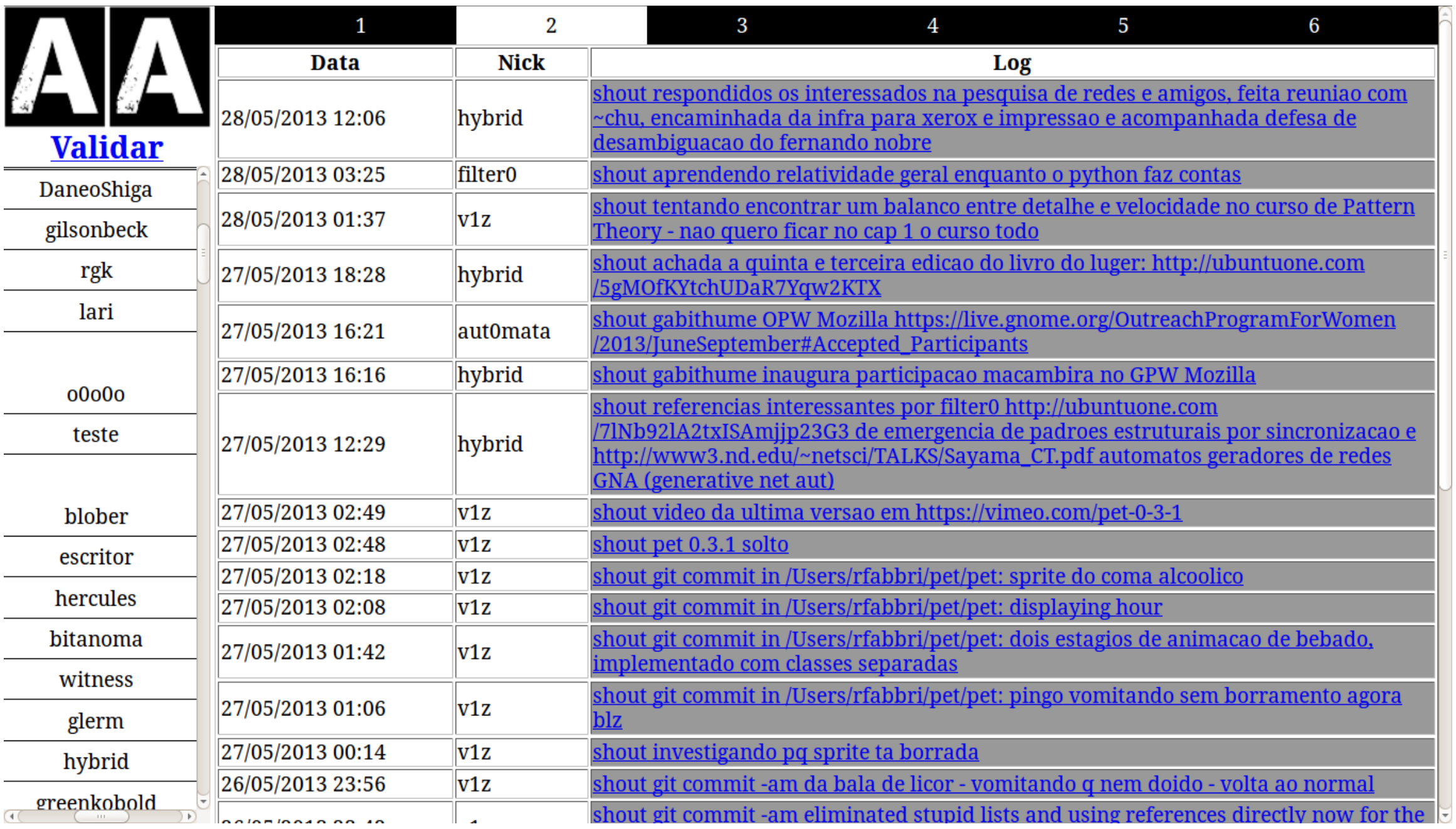

| Data | Nick | Log |
|---|---|---|
| 28/05/2013 12:06 | hybrid | shout respondidos os interessados na pesquisa de redes e amigos, feita reuniao com ~chu, encaminhada da infra para xerox e impressao e acompanhada defesa de desambiguacao do fernando nobre |
| 28/05/2013 03:25 | filter0 | shout aprendendo relatividade geral enquanto o python faz contas |
| 28/05/2013 01:37 | v1z | shout tentando encontrar um balanco entre detalhe e velocidade no curso de Pattern Theory - nao quero ficar no cap 1 o curso todo |
| 27/05/2013 18:28 | hybrid | shout achada a quinta e terceira edicao do livro do luger: http://ubuntuone.com /5gMOfKYtchUDaR7Yqw2KTX |
| 27/05/2013 16:21 | aut0mata | shout gabithume OPW Mozilla https://live.gnome.org/OutreachProgramForWomen /2013/JuneSeptember#Accepted_Participants |
| 27/05/2013 16:16 | hybrid | shout gabithume inaugura participacao macambira no GPW Mozilla |
| 27/05/2013 12:29 | hybrid | shout referencias interessantes por filter0 http://ubuntuone.com /7lNb92lA2txISAmjjp23G3 de emergencia de padroes estruturais por sincronizacao e http://www3.nd.edu/~netsci/TALKS/Sayama_CT.pdf automatos geradores de redes GNA (generative net aut) |
| 27/05/2013 02:49 | v1z | shout video da ultima versao em https://vimeo.com/pet-0-3-1 |
| 27/05/2013 02:48 | v1z | shout pet 0.3.1 solto |
| 27/05/2013 02:18 | v1z | shout git commit in /Users/rfabbri/pet/pet: sprite do coma alcoolico |
| 27/05/2013 02:08 | v1z | shout git commit in /Users/rfabbri/pet/pet: displaying hour |
| 27/05/2013 01:42 | v1z | shout git commit in /Users/rfabbri/pet/pet: dois estagios de animacao de bebado, implementado com classes separadas |
| 27/05/2013 01:06 | v1z | shout git commit in /Users/rfabbri/pet/pet: pingo vomitando sem borramento agora blz |
| 27/05/2013 00:14 | v1z | shout investigando pq sprite ta borrada |
| 26/05/2013 23:56 | v1z | shout git commit -am da bala de licor - vomitando q nem doido - volta ao normal |
| [illegible] | [illegible] | shout git commit -am eliminated stupid lists and using references directly now for the |

Figure 2: The AA Report Aggregator V. 0.1 front page displaying the latest AA messages of users hybrid, filter0, v1z and aut0mata on distributed activities for a range of globally collaborating entities (LabMacambira.sf.net, IPRJ/UERJ, IFT/UNESP, IFSC/USP, OPW/Mozilla, Pula Pirata Comics, Inc.). Each message is an AA shout which, when grouped, constitutes an AA session.
Source: the authors.

## 3.3 PEER VALIDATION

No set bosses or leaders are required in an ideal application of the AA methodology. In practice, the need for centralized administrative overhead is greatly reduced and made flexible due to the self-regulating mechanisms of the approach. Hence the name 'Algorithmic Auto regulation' and other implicit interpretations of the AA acronym and logo.

The primary mechanism to achieve decentralization in AA is peer-to-peer coordination by harnessing social behavior. It can be deliberate or implicit. In order to prevent spamming and to improve the overall quality of AA reports, each AA session must be validated by another developer. More specifically, all reports are read by someone that will mark them collectively as 'valid' or 'invalid' and may optionally write comments about the specific session and quality of micrologs. The developer in charge of validating any given session is randomly assigned by the AA web server, which sends out an e-mail to the developer with an URL to a validation interface.

Peer validation also helps in making decentralized collaboration more cohesive by encouraging members to be minimally aware of peer activities, even when these are not immediately useful for accomplishing the task at hand. We have observed that decentralized teamwork can get so efficient at actual production that the team gets short-sighted in terms of coordination: non-communicating subteams can get formed if care is not taken, causing a fragmentation of the collective. Peer validation is one way to help avoid fragmentation and is an essential mechanism of decentralized team auto regulation.

## 4 RESULTS AND DISCUSSION

Easy and effective GSD team management is the main purpose of the AA methodology. To verify its success, the proposed methodology was employed by the Lab Macambira free software development collective to auto-regulate a group of nine developers in July-December 2011, three of which are coauthors of the present work. The goals of the team was to work on an array of strategic free software projects in audiovisual and Web technology, contributing directly to official development, submitting bug patches or committing new features to source code[2]. Of particular interest are the strategies adopted to tailor AA to a real context, which are useful as general guidelines.

The team members had different levels of experience on software engineering for large and distributed free software projects like Scilab and Mozilla. To level the field, one month of training was conducted by three experienced developers (the first authors of the present work), starting by teaching infrastructure tools like bug trackers, required programming languages, version control systems, and build systems. After this initial period, a starter project was proposed for new developers: to submit a bug fix or implement a new feature for a large free software project and have an accepted patch or commit to the official repository by the end of July.

Developers passing the starter project would be deemed 'initiated' and called a 'Macambira' developer, and were hired for paid work using the AA methodology for the remainder of the semester. To illustrate the breadth of the resulting contributions, Table 1 summarizes the effective accepted commits of each successfully initiated developer to free software projects in 2011, which used the AA methodology. The first column lists applications to which contributions were officially accepted and whose development process was tracked and publicly documented using AA. The second column shows the pseudonym of the committers (at Lab Macambira it is common practice to employ pseudonyms in AA in order to enhance privacy).

In one month, each developer officially contributed to one or many free software projects. Many developers started the initiation training with no knowledge of what free software was and ended that period becoming a free software developer. During that month, the same team of trainees also developed the first version of the AA system and used AA auto-regulate in their activities, even while developing other aforementioned free software projects. Thus, AA and the associated software system was tested, prototyped, and developed in close contact with actual practice. The source code of AA — both the clients that send micrologs and the AA Web server — is publicly available as free software[3]. Moreover, the entire AA log data of the Lab Macambira team from 2011 to the present time is also available on-line, which, together with public Git logs from each project, documents the claims in Table 1 and enables further analysis of the corresponding data.

[2] LabMacambira.sf.net: http://labmacambira.sourceforge.net.

Table 1: Free and open software projects that received documented contributions from successfully initiated developers or ‘Macambiras’ using the AA methodology.

| **Application** | **Commiters** |
|---|---|
| Mozilla Firefox | daneoshiga, bzum |
| Evince | hick209, bzum, marcicano, mquasar |
| BePDF / Xpdf | marcicano |
| Ekiga | flecha |
| Empathy | fefo |
| Lib Folks (Telepathy) | kamiarc |
| Scilab | v1z, humannoise |
| VxL | v1z |
| ImageMagick | v1z |
| OpenOffice | hick209 |
| Puredata | v1z, automata, greenkobold, gilson, bzum |
| Puredata OpenCV | v1z |
| Puredata GEM | v1z, fefo, hick209 |
| Puredata PDP | v1z, fefo, hick209 |
| ChucK | rfabbri, automata |
| ChucK MiniAudicle | rfabbri, automata |
| Mozilla Firefox WebRTC | automata |
| OSC-Web | automata |
| Live-Processing | automata |
| Chuck-Wiimote | automata |
| Audiolet | automata |
| Extempore | automata |

Source: the authors

After the initial training period of one month, the initiated ‘Macambiras’ worked during 6 additional months on a large range of free software projects, divided into work groups — each work group focusing on a specific theme like video, audio and web. Funding sources were mainly contracts, freelance, and the direct support of the Pontão Nós Digitais NGO. The team also created a range of completely new free software applications, as listed in Table 2[4]. It is interesting to note the heterogeneity of projects and their areas of application.

Table 2: Software projects created by Lab Macambira since July 2011 using the AA methodology, together with a short description and the technologies involved.

[3] AA client and server source code available at: http://wiki.nosdigitais.teia.org.br/AA; AA logs: www.pulapirata.com/skills/aa

[4] http://wiki.nosdigitais.teia.org.br/LabMacambira.

| Application | Description | Technologies |
|---|---|---|
| AA | Algorithmic Auto regulation | Python, PHP |
| Ágora Communs | System for online deliberations | PHP |
| SIP | Scilab Image Processing toolbox | C, Scilab |
| Animal | An Imaging Library | C |
| TeDi | Test Framework for Distance Transform Algorithms | C, Shell, Scilab |
| Macambot | Multi-use IRC Bot | Python |
| Conferência permanente | Platform for the permanent conference of the rights of minors | PHP, JavaScript |
| CPC | Center for the Brazilian culture representation groups | Python, Django |
| Timeline | Interactive timelines on the Web | JavaScript |
| Imagemap | Interactive labeling of on-line photos | JavaScript |
| ABT | Program for real-time sound execution and musical rhythmic analysis | Python |
| EKP | Emotional Kernel Panic | Python, ChucK |
| SOS | Aggregation and diffusion of popular and native knowledge about health | Python, Django |
| Creative Economy | Platform for creative collaborative solidary economy of culture hubs and entities | Python, Django |
| OpenID Integration | Adaptations to existing software for unified login through OpenID | PHP |
| pAAnel | Dashboard for the real-time visualization of Lab Macambira activity | Python, Django |
| Georef | Collection of scripts to be used as reference and a GIS platform to map public data of interest to citizens | Python, Django |
| AirHackTable | Software system to generate sound by real-time 3D tracking of flying objects | Puredata, C++, Scilab |

Source: the authors

While using the AA system, developers learned to work asynchronously with others and got used to the habit of periodically updating their status on their projects. Each programmer was given the chance to work with considerable freedom, in any place and time of preference. The strictest required responsibility was that of using AA for at least one 2h session per day, while working on the agreed tasks. The online pAAnel allowed each developer to quickly grasp activities of interest from others while avoiding interrupting them, a process further aided by the screencasts.

Adjustments to the task deadlines and milestones (which were managed in Trac and, more recently, on Github) were performed based on observed progress of individuals and of the labMacambira.sf.net team as a whole. The numerous inexperienced newcomers benefited from a friendly environment suitable for fast learning by the use of AA as a flexible and simple transparency system. This was key for the motivation and fixation of new contributors, which is a central issue in free software development (SANTOS *et al.*, 2013). Updates from the team were transmitted not on a person-to-person basis, but rather on a person-to-team basis through the available online progress information.

As of the time of this writing, Lab Macambira comprises over fifteen software developers, with the logs registering over eight man-years of work using AA. Similar team statistics from AA logs are continuously updated and displayed in graphical widgets as part of a customized version of the pAAnel dashboard (Figure 3). Key developers among those trained in 2011 continue to work in the collective as volunteers and contracted developers with foundations like Mozilla.

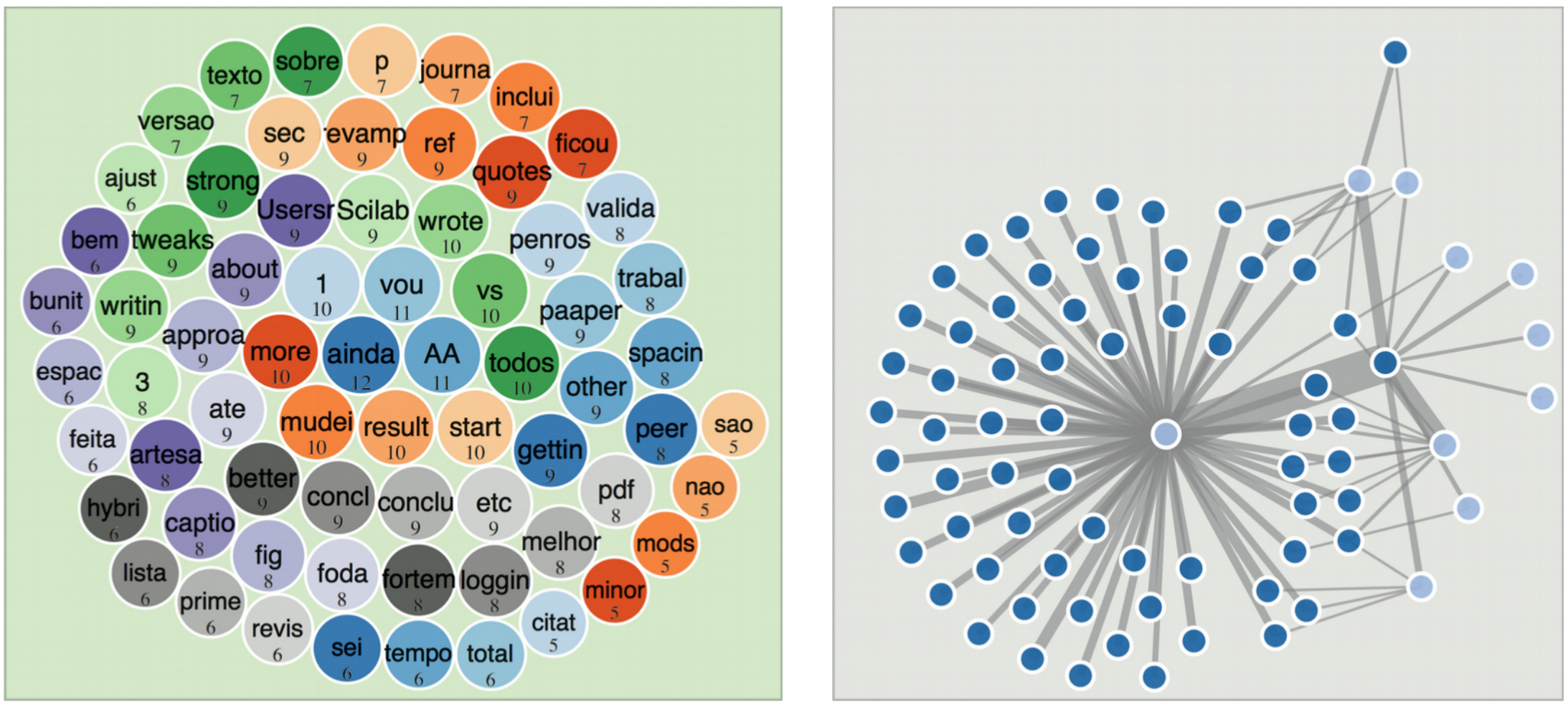

Figure 3: Visualizations of team dynamics from AA logs. *Left:* a bubble word cloud of latest messages reflecting emerging concepts; *Right:* bipartite graph linking developers (light blue) to frequently used terms (dark blue), reflecting the formation of communities. In the AA instance of Lab Macambira, these graphs are interactive and continuously updated from a window of 125 shouts.
Source: the authors

## 5 CONCLUSIONS

In a scenario where Global Software Development is growing across the entire software industry, there is an increasing need for methodologies to deal with its potential disadvantages while amplifying its powerful advantages.

This paper has presented the highly scalable AA methodology, designed to connect a series of large or small groups of software developers working from different countries or in the same room. The methodology is built around a simple system where each developer takes note of his work by posting a periodic log of short text sentences or micrologs. The sum of those activity logs, along with an entire session of work, results in a complete unit of report. The report is made publicly available through a Website and is validated by peers that are randomly selected by the AA Web server. Real-world data on practical experience with this approach has been collected and reported, involving a team of paid free software developers, Lab Macambira, which since 2011 has developed new free and open source software for a vast number of high-end applications using AA.

AA is not limited to a work-management tool, but acts as a self-regulation methodology to improve the temporal sensitivity and sensibility of individuals, helping divide complex tasks in time into small chunks or sessions, and also reducing the need for extensive reports or unnecessary meetings. By asking users to publish minimal text sentences as a continuous log feed, the methodology avoids disturbing the flow of developers which are heavily concentrated in programming. These developers just have to type a few task-related words and go back to coding; others get updated as needed for their task. Interruption-free communication is achieved – AA interruptions are in context, decentralized and maximally useful.

The AA methodology is not restricted to software development. As of this writing there is an entertainment studio, Pula Pirata that has been using AA to manage their creative activities[5]. Other people with no software background, like social scientists, musicians and activists also have been using AA, contributing to its broader improvement[6].

There are many aspects of this work to be further explored. Additional ubiquitous client interfaces for micrologging from different existing tools beyond IRC and Twitter, e.g., other web social services and e-mail, would greatly make the use of AA easier and more widespread, turning it into a truly replicable system. Another research direction is to analyze the actual work logs generated by the Lab Macambira and Pula Pirata collectives since 2011 to recognize behavioral patterns in individuals and their creative process. It would also be desirable to carry out more specific experiments by harnessing recent research on the psychology of time responses (CAETANO *et al.*, 2012; GUILHARDI, *et al.* 2010). This would enable the scientific testing of the claims made in this paper to refine the methodology, its mechanisms and parameters.

## 6 ACKNOWLEDGMENTS

Authors acknowledge the financial support from Pontão Nós Digitais, and Ricardo Fabbri acknowledges support from FAPERJ/Brazil 111.852/2012. Authors also thank AA: the present research project and even this manuscript were written using AA. The complete log for this paper is on-line at www.pulapirata.com/skills/aa. Finally, authors are also grateful to IFSC/USP, IPRJ/UERJ, IFT/UNESP and all AA users and collaborators, especially those who coded AA hacks for logging through shell and bots, and those who coded the different Web interfaces in use.

5 http://www.pulapirata.com.

6 A small representative sample of our public logs reveals the pseudonyms of **activists** (flecha, humanoise, angelina), **social scientists** (humanoise), **musicians** (audiohack, glerm, cravelho), and **architects** (prestoppc), who have started to use AA due to convergences with digital media technology at Lab Macambira.